\documentclass[a4paper,10pt]{article}
\pdfoutput=1
\usepackage{jcappub}
\usepackage{mathtools}
\usepackage{ulem}
\usepackage{bm}
\usepackage{xcolor}
\usepackage{subcaption}
\usepackage{float}
\usepackage{orcidlink}
\usepackage{xcolor}
\usepackage{marginnote}

\renewcommand{\thesection}{\arabic{section}}

\def\cob{\color{blue}}

\newcommand{\au}[2]{#1.~#2}

\newcommand{\oarX}[1]{\href{http://arxiv.org/abs/#1}{{\ttfamily\cob arXiv:#1}}}
\newcommand{\arX}[1]{\href{http://arxiv.org/abs/#1}{{\ttfamily\cob arXiv:#1}}}
\newcommand{\doin}[6]{\href{http://dx.doi.org/#1}{{\cob {\it #2} {\bf #3 #4} (#6) #5}}}
\newcommand{\doinn}[5]{\href{http://dx.doi.org/#1}{{\cob {\it #2} {\bf #3} (#5) #4}}}
\newcommand{\doij}[5]{\href{http://dx.doi.org/#1}{{\cob {\it #2} {\bf #3} (#5) #4}}}

\newcommand{\tia}[1]{\textit{#1},}

\def\laq{~\raise 0.4ex\hbox{$<$}\kern -0.8em\lower 0.62ex\hbox{$\sim$}~}
\def\gaq{~\raise 0.4ex\hbox{$>$}\kern -0.7em\lower 0.62ex\hbox{$\sim$}~}

\def\beq{\begin{equation}}
\def\eeq{\end{equation}}
\def\bea{\begin{eqnarray}}
\def\eea{\end{eqnarray}}

\def \fp {{\dot{\phi}}}

\def \pa {\partial}
\def \ra {\rightarrow}
\def \ti {\widetilde}
\def \la {\lambda}
\def \ls {\lambda_{\rm s}}

\def \Mp {M_{\rm P}}

\def \Da {\Delta}
\def \da {\delta}
\def \b {\beta}
\def \a {\alpha}
\def \ap {\alpha^{\prime}}

\def \ga {\gamma}
\def \sg {\sigma}

\def \da {\delta}
\def \ep {\epsilon}
\def \r {\rho}

\def \Om {\Omega}
\def \noi {\noindent}

\def \fp {\dot{\phi}}
\def \bp {\dot{\beta}}
\def \gp {\dot{\gamma}}
\def \fb {\overline \phi}

\def \bp {\dot{\beta}}

\title{A simple example of ``non-minimal" Pre-Big Bang scenario }

\author[a,b]{P. Conzinu\,\orcidlink{0000-0002-7290-7790},}
\author[c]{M. Gasperini\,\orcidlink{0000-0001-9117-8303}}
\author[c]{and E. Pavone\,\orcidlink{0000-0002-3022-4545}}

\affiliation[a]{Scuola Superiore Meridionale, Via Mezzocannone 4, 80138 Napoli, Italy}

\affiliation[b]{
Istituto Nazionale di Fisica Nucleare (INFN), Gruppo Collegato di Parma, Parco Area delle Scienze 7/A, I-43124, Parma, Italy
}

\affiliation[c]{
Dipartimento di Fisica, Universit\`a di Bari, 
Via G. Amendola 173, 70126 Bari, Italy,\\
and Istituto Nazionale di Fisica Nucleare, Sezione di Bari, Italy
}

\emailAdd{p.conzinu@ssmeridionale.it}
\emailAdd{gasperini@ba.infn.it}
\emailAdd{eliseo.pavone@ba.infn.it}

\abstract{We give an example of non-minimal pre-big bang scenario able to produce the PTA signal considering a modified evolution of the high-curvature string phase, including the contribution of high-energy string sources. We use a fluid-dynamic model of sources and show that their effective viscosity breaks the $S$-duality symmetry of the tensor-axion perturbation spectra, as in general expected for the non-minimal scenario. }

\keywords{Primordial gravitational waves (Theory), String cosmology, Pre-big bang
 
\vskip18pt 

\noindent{\bfseries\large\sffamily{Preprints:}}~BA-TH/810-25} 

\begin{document}
\maketitle

\section{Introduction}  
\label{sec1}

In a recent paper \cite{1} we have discussed the possible interpretation of the signal detected by the multiple Pulsar Timing Array (IPTA) collaborations, including NANOGrav \cite{2,2a}, the Parkes PTA (PPTA) \cite{2b,2c}, the European PTA (EPTA) in partnership with the Indian PTA (InPTA) \cite{2d,2e}, and the Chinese PTA (CPTA)\cite{2f}, as due to a stochastic background of relic primordial gravity-wave (GW) radiation produced in the context of the Pre-Big Bang  (PBB) scenario  \cite{3,3a,4,Lidsey:1999mc, 5}, based on the duality symmetries of the string cosmology equations \cite{6, Meissner:1991zj, Meissner:1991ge, Gasperini:1991ak, Sen:1991zi}.

Such an interpretation, as discussed in \cite{1}, is impossible for the GW spectrum produced by the so-called ``minimal" version of such a scenario (see e.g. \cite{18,7}), as it requires appropriate modification of the GW spectrum amplified during the high-curvature string phase. 
The aim of this paper is to provide a simple but physically motivated example of how such modifications could be implemented in the context of a ``non-minimal" model of PBB evolution still described by the standard string cosmology equations. 

Let us start by recalling the basic elements of the minimal PBB scenario \cite{3,3a,4}: the cosmological evolution starts asymptotically from the string perturbative vacuum and, after a low-energy phase of growing curvature and growing dilaton, reaches a high-energy phase with string-scale curvature, preceding the bounce and the beginning of the standard, decreasing curvature, frozen dilaton, post-big bang evolution. The high-energy string phase of the minimal scenario, in particular, is described by an epoch in which the curvature of the internal and external dimensions is nearly constant at the string scale, and the effective string coupling, controlling the string loop corrections, is growing (according to the explicit model first discussed in \cite{20}). 

Let us try to preserve the above-mentioned properties of the string phase also in the modified version of the non-minimal scenario. To this purpose we recall that the string phase of the minimal scenario is described by a fixed-point solution of the {\it vacuum} gravi-dilaton equations, including higher-curvature corrections to first order in the string $\ap$ expansion \cite{20}. Therefore, we will modify such a scenario by only adding to the background dynamics the contribution of non trivial (high-energy) effective matter sources; and we will ask whether, depending on the properties of these  sources, the amplified tensor and axion perturbations may be characterised by  spectral powers different from those of the minimal model, so as to satisfy the constraints discussed in \cite{1} needed to  produce the signal detected by the IPTA collaboration. Finally, we will describe the matter sources as a higher-dimensional fluid (as usual in the context of homogeneous cosmologies), with possibly anisotropic pressure (in case of a different dynamics for internal and external spatial dimensions), and with possibly intrinsic shear viscosity (to produce a scenario with broken $S$-duality symmetry \cite{12}).

We have found, with explicit calculations, that there are various models of sources able to produce the required scenario. For the illustrative purpose of this paper, and for the sake of brevity, we shall mainly concentrate our discussion on three cases: $i)$ radiation, described by fluid sources with traceless stress tensor; $ii)$ gas of primordial unstable strings \cite{Gasperini:1990xg, Gasperini:1991rv}, described by fluid sources with equation of state, in an isotropic $D$-dimensional geometry, given by $p/\rho=-1/(D-1)$; $iii)$ gas of string holes (i.e. string-size black holes) \cite{23a,23b,23c,Bitnaya:2023vda,Conzinu:2024bdf},  described by fluid sources whose pressure $p$ is related to the dilaton charge $\sg$, in the string frame ($S$-frame), by $p= \sg/2$. We have chosen these three examples because we may naturally expect the presence of these type of sources in the high-curvature regime of the string phase. 

Finally, let us anticipate here a result which, in our opinion, is probably one of the most interesting ones of this paper: the new allowed background solutions describing the string phase, and satisfying the required constrains needed for a successful non-minimal scenario,  may be characterised by  spatial dimensions evolving with an effective Hubble parameter of opposite sign to that of the initial asymptotic regime. For instance, by an external space which is contracting at constant curvature (or even flat), instead of being expanding. Note that this is not  an unphysical result (as it might seem at first glance) but, on the contrary, an interesting (and in principle expected) result, since it describes just the same type of  kinematics of the high-curvature phase obtained in the context of regular and self-dual string models of bounces (see e.g. 
\cite{cin,Gasperini:2023tus,Conzinu:2023fth}). 

The paper is organised as follows. In Sect. \ref{sec2} we introduce the background dynamics and its perturbations for our model of high-curvature, non-vacuum string phase. In Sect. \ref{sec3} we present and discuss, for the chosen examples of fluid sources, their effects on the dynamics of the string phase, the resulting modification of the perturbation spectra,and their behaviour in the spectral plane to be compatible with the PTA signal.
We also consider examples of viscous sources, producing non-minimal models with broken $S$-duality symmetry \cite{12}. Sec. \ref{sec4} is devoted to our concluding remarks. The explicit form of the modified background equations is reported in App. \ref{secA}, and the canonical evolution of axion perturbations, including first-order, higher-curvature corrections, is presented in App. \ref{secB}. The main details of the non-minimal GW spectrum, and the theoretical, phenomenological and self-consistency constraints to be imposed for its compatibility with the detected PTA signal,  are finally summarised in App. \ref{secC}. 

\section{A non-vacuum string phase: background and perturbation equations} 
\label{sec2}

Let us start with the $S$-frame action used to describe the string phase in the context of the minimal PBB scenario \cite{20}, and obtained to first order in  $\ap$ from the two-loop sigma model through a field redefinition (see e.g. \cite{Tsey}) which gives an action without higher-than-second derivatives in the corresponding equations of motion. By including the string antisymmetric tensor\footnote{This background field is an essential ingredient for a  phenomenologically complete scenario able to produce the today observed spectrum of scalar metric perturbations \cite{12}.} 
$B_{\mu\nu}=-B_{\nu\mu}$, and adding the source contribution described by the matter action $S_m$,  the total action, in a general $D$-dimensional space-time geometry, is then given by
\bea
S= &&S_m -{1\over 2\ls^{D-1}} \int d^Dx \sqrt{-g} e^{-\phi} \left[R+ \pa\phi^2 - {1\over 12} H_{ABC}^2 - \ap\a_0\left(R_{GB}^2 - \nabla \phi^4\right)
 \right.
\nonumber \\ &&
\left.
-\ap a_0  H_{ABC}^2\left({1\over 12} g^{MN} R_{MN} + {5\over 4} \nabla \phi^2 \right) \right.
-\ap a_0 H^{MAB} H_{NAB} \left( R_M\,^N -\nabla_M \phi \nabla^N \phi \right)
\nonumber \\ &&
\left. 
- \ap a_0 \left({1\over 24} H_{MNL}H^N\,_{RA}H_S\,^{MA}H^{RSL}-
{1\over 8} H_{MRL}H_N\,^{RL}H^{MSA}H^N\,_{SA}\right) \right.
\nonumber \\ &&
\left. 
+{1\over 2} \ap a_0 R_{ABCD} H^{ABM}H^{CD}\,_M \right].
\label{21}
\eea
Here $\ls^2\equiv 2\pi \ap$ is the string length parameter, $\phi$ is the dilaton, $H_{ABC}= \pa_A B_{BC} +  \pa_B B_{CA}+\pa_C B_{AB}$, where $B$ is the NS-NS two-form (related in four dimension to the so-called Kalb-Ramond axion), and $a_0$ is a numerical parameter depending on the given type of string model: in particular, $a_0=1/4$ for the bosonic string, and $a_0=1/8$ for the {heterotic} superstring. Finally, $R_{GB}^2$ is the Gauss-Bonnet quadratic curvature invariant, and capital latin indices run from $0$ to $D-1$.  


\subsection{Background dynamics of the string phase}
\label{sec21}

Consider a higher-dimensional geometry of Bianchi-I type, describing the product of two isotropic subspaces with $d=3$ (external) and $n$ (internal) dimensions, represented (using the cosmic time $t$) by the metric
\bea
&&
g_{AB}= {\rm diag} \,(g_{\mu\nu}, \ga_{mn})=  {\rm diag} \,(1, -a^2 \da_{ij},   -b^2 \da_{mn} ), ~~~~~~~~~ a= e^{\b(t)}, ~~~ b = e^{\ga(t)},
\label{22}
\eea
where Greek indices run from $0$ to $3$, Latin indices $i,j$ from $1$ to $3$,
Latin indices $m,n$ from $4$ to $4+n$, capital Latin indices from $0$ to $D-1=3+n$. 

By assuming that the background value of the Kalb-Ramond strength tensor is vanishing ($H_{ABC}=0$) we find, for the given class of geometric backgrounds, that the action (\ref{21}) reduces to the following effective action for the variables $\{N,\phi, \b, \ga\}$:
\bea
S =&& S_m -{1\over 2\ls^{D-1}}  \int  {dt\over N}e^{3\b + n \ga -\phi} \left\{\left[ \fp^2+ 6 \bp^2 +n(n-1)\gp^2+ 6 n \bp \gp - 6 \bp\fp-2n \gp \fp\right]\right.
\nonumber \\ &&
\left.
-{\ap a_0\over N^3}\left[ c_2 \gp^4+ c_3\fp\bp^3 + c_4 \gp^3(\fp-3 \bp)- \fp^4 + c_5\bp\gp^2(\fp-\bp)+c_3 n \gp \bp^2(3 \fp- \bp)\right]
\right\},
\nonumber \\ 
\label{23}
\eea
where $N^2= g_{00}$ is the so-called lapse function, and where:
\beq
c_2 = -{n\over 3}(n-1)(n-2)(n-3),~~~~~~~
c_4= {4n\over 3}(n-1)(n-2), ~~~~~~~ c_5= 12 n (n-1)
\label{24}
\eeq
(we are using the same notations as in \cite{20}). 

As already mentioned we shall use a fluid model of sources, possibly coupled to the dilaton, with possibly different external ($p$) and internal ($q$) pressure (in agreement with the spatial anisotropy of the metric (\ref{22})), and with the possible presence of intrinsic shear viscosity 
$\eta_V$. The variation of the matter action with respect to the metric and to the dilaton then gives the so-called dilaton charge $\sg$ and the canonical stress tensor, defined as usual by
\beq
{\sg \over 2}=- {1\over \sqrt{-g} }{\da S_m \over \da \phi}, ~~~~~~~~~~~~~~~~
T^{AB}= -{2\over \sqrt{-g}} {\da S_m \over \da g_{AB}}.
\label{24a}
\eeq
The stress tensor, taking into account the particular spatial structure of the metric (\ref{22}) describing the direct product of  $3$-d and  $n$-d isotropic subspaces, and including (in the standard form) the shear viscosity contribution, can be covariantly written in general as follows:
\beq
T_{AB}= \left(\rho+ \ti p\right) u_A u_B - \ti p g_{AB}+\left( \ti q - \ti p \right) v_A v_B 
 - \eta_V \left[u_{(A|} u^M \nabla_M u_{|B)} - \nabla_{(A} u_{B)} \right] 
\label{24b}
\eeq
(round brackets denote symmetrisation). 
Here 
$u^A$ and $v^A$ are, respectively, time-like and space-like vectors satisfying the conditions $u_A u^A=1$, $v_A  v^A=-n$ (see e.g. \cite{Letelier:1980mxb, Herrera:1997plx} for the formal description of anisotropic fluid sources in the special case of $D=1+3$ dimensions). Finally, the tilde symbol over the internal and external pressures denotes as usual the possible presence of the shear viscosity contribution $\eta_V$ as follows:
\beq
\ti p = p+{2\over 3+n} \eta_V \nabla_A u^A, ~~~~~~~~~~~~~
\ti q =q+{2\over 3+n} \eta_V \nabla_A u^A.
\label{24c}
\eeq
An explicit computation for the background metric of Eq. (\ref{22}), performed in the comoving gauge where  $u_A u^B= \da_A^0 \,\da _0^B$ and 
$v_A v^B =- \da_A^m \,\da _n^B \,\da_m^n$, 
then gives the following modified components of the source stress tensor:
\beq
T_0^0= \rho, ~~~~~~~
T_i^j= - \da_i^j \left[p+{2 \eta_V \over 3+n} n (\gp-\bp)\right], ~~~~~~~
T_m^n=-\da _m^n \left[q+ {2 \eta_V \over 3+n} 3(\bp-\gp) \right] 
\label{24d}
\eeq
(note that in the limit of an isotropic geometry, $\bp=\gp$, there are no contributions of the shear viscosity to the stress tensor and to the background equations, as expected \cite{Dolgov}). 
We shall assume, in the following, that the non-viscous components of the source satisfy the perfect fluid equation of state, namely
namely:
\beq
p=w_1 \rho, ~~~~~~~~~~~~~q= w_2 \rho, ~~~~~~~~~~~~~
{\sg \over 2} = w_3 \rho,
\label{25}
\eeq
with $w_i=$ const. 

The explicit equations governing the dynamics of the string phase are now obtained by varying the gravi-dilaton part of the action (\ref{23}) with respect $N, \b, \ga, \phi$, and by adding the associated contribution of the fluid sources. After the variation we set $N=1$ (cosmic time gauge), and we look, as in the minimal model, for background solutions with constant curvature and with a dilaton which is linearly evolving with respect to the cosmic time coordinate. Namely, we impose $\dot a /a= \bp=$ const, $\dot b/b= \gp=$ const, $\fp=$ const, where the dot denotes cosmic time derivative. The explicit form of the resulting equations, including the fluid source contribution parametrised as in Eqs. (\ref{24d}), (\ref{25}), is reported in Appendix \ref{secA}. It can be easily checked that the sources, to be consistent with the given type of background solutions, must satisfy the conditions
\beq
\ls^{D-1} e^\phi \rho\equiv C = {\rm const}, ~~~~~~~~~~~~~
\la_s^{D-2} \eta_V \,e^\phi \equiv H_V = {\rm cost}.
\label{26}
\eeq
(see Eqs. (\ref{a1})--(\ref{a4})). 
It should be stressed, also, that the introduced high-energy fluid sources, in order to be relevant for the kinematics of the string-phase background and for the corresponding perturbation equations, are assumed to provide contributions of the same order as the higher-curvature $\ap$ corrections. Hence the constant parameter $C$, as well as as the viscosity parameter $H_V$, quite independently of their possible microphysical origin and of the  considered order of the $\ap$ expansion, are expected to have a typical order of magnitude controlled by the string scale, and will be treated as phenomenological parameters of the high-energy stringy regime.

The equations reported in Appendix \ref{secA} are a system of four algebraic equations for the  unknown constants $\bp,\gp,\fp,C,H_V, w_1,w_2,w_3$, and we shall look for non trivial solutions for appropriate fluid models. In the absence of sources only three of the above equations are independent and, in the particular limit of an isotropic geometry, $\bp=\gp$, one can recover the known solutions of the minimal model presented in \cite{20}. In that case, however, we cannot satisfy the conditions required to produce a GW spectrum compatible with the PTA signal. 

For an explicit formulation of the required conditions let us now introduce a few details on the evolution of axion and tensor perturbations in the chosen model of string-phase background. 


\subsection{Axion perturbations with $\ap$ corrections}
\label{sec22}

Let us report here the basic results needed to compute the axion spectrum (which is a crucial ingredient, for the PPB scenario, to produce the today observed background of isocurvature scalar metric perturbations via the curvaton mechanism \cite{12,13,14,15}). 

We will assume that the matter sources are not directly coupled to the Kalb-Ramond field, so that the resulting axion-perturbation equation will be exactly the same as that of the minimal scenario. However, we will explicitly take  into account the possible contribution of the $\ap$ corrections typical of the string phase\footnote{This is a point which, to the best of our knowledge, has never been taken into account for the axion in previous papers.}, and present indeed in the action (\ref{21}). 

Let us start with the action (\ref{21}), consider the first order perturbations of the Kalb-Ramond field strength, $H_{ABC} \ra H_{ABC}+ \da H_{ABC}$, assume that the zeroth-order background contribution is vanishing, $H_{ABC}=0$, and compute from (\ref{21}) the effective action quadratic in the perturbation $\da H_{ABC}$. We are interested, in particular, in the dynamics of the pseudo-scalar Kalb-Ramond axion field $\chi$, i.e. 
the spacetime dual of the  four-dimensional component of the Kalb-Ramond perturbations. It is defined by
\beq
\da H^{\mu\nu\a} = {e^\phi\over \sqrt{|\ga|}} {\epsilon^{\mu\nu\a\b}\over \sqrt{-g}} \nabla_\b \chi \equiv e^{\fb} \eta^{\mu\nu\a\b} \pa_\b \chi,
\label{27}
\eeq
where $\mu, \nu, ...=0,1,2,3$, $\eta^{\mu\nu\a\b}$ is the four-dimensional totally antisymmetric tensor, and  $e^{\fb}= e^\phi/\sqrt{|\ga|} = e^\phi/b^n \equiv g_4^2$ is the square of the effective four-dimensional string coupling.

We shall now perturb the action (\ref{21}) according to Eq. (\ref{27}) up to terms of quadratic order $(\pa \chi)^2$ and, by taking into account that $\chi= \chi(t, x_i)$, we shall use the Fourier component of the axion perturbation $\chi_k$, such that $\nabla^2 \chi_k= \pa_i\pa^i \chi_k =- k^2 \chi_k$. By introducing the conformal-time coordinate $\eta$ such that $dt= a d\eta$, and factorising the volume integral  $\int d^ny$ over the internal spatial dimensions, we then obtain the following quadratic perturbed action for $\chi_k$:
\beq
\da_H S= -{1\over 2 \ls^2} \int d \eta \left[z^2(\eta) \chi_k^{\prime 2}-
k^2 \chi_k^2 y^2(\eta)\right].
\label{28}
\eeq
Here a prime denotes differentiation with respect to the conformal time, and $z$ and $y$ are functions of time depending on the background fields (see Appendix \ref{secB} for an explicit computation and a general definition of these functions). From the action (\ref{28}) we can immediately obtain the equation of motion for the axion perturbation $\chi_k$:
\beq
\chi'' + 2 {z'\over z} \chi' + k^2 {y^2\over z^2}\, \chi =0, 
\label{29}
\eeq
and the corresponding evolution equation for the canonical variable $v_k= z \chi_k$ which diagonalises the kinetic term of the action (\ref{28}):
\beq
v_k'' + k^2 v_k - V_k(\eta) v_k =0, ~~~~~~~~~~~~
V_k (\eta)= {z''\over z} -{k^2\over z^2}\left(y^2 -z^2\right).
\label{210}
\eeq

Let us now take into account that our background describing the string phase is characterised by constant values of the parameters $\bp, \gp, \fp$ controlling the curvature and the dilaton dynamics. In such a case it turns out that the two functions $z$ and $y$ are proportional (see their explicit expression in Appendix \ref{secB}), and are given by
\beq
z= \a_1 \xi_\sg, ~~~~~~~~~~
y= \a_2 \xi_\sg, ~~~~~~~~~~~~~~
\xi_\sg = a \,b^{-n/2} e^{\phi/2},
\label{211}
\eeq
where $\a_1,\a_2$ are numerical constants of order one. The canonical equation (\ref{210}) determining the axion spectrum thus reduces to
\beq
v''_k + \left(k^2c_s^2 - {\xi_\sg''\over\xi_\sg}\right) v_k=0,
\label{212}
\eeq
where $\xi_\sg$ is the so-called axion pump field, and $c_s = \a_2/\a_1\sim 1$ is the effective ``sound velocity" of the axion fluctuations (see e.g. \cite{Con,Conzinu:2023fui} for the possible physical consequences of a speed $c_s \not=1$). 

It should be noted that the previous result has been explicitly obtained from the action (\ref{21}), which includes higher-curvature corrections truncated to first-order in $\ap$. However, for the particular case of the string-phase dynamics  describing constant curvature and constant  kinetic energy of the dilaton field, also  higher-order $\ap$ corrections give similar contributions to the axion perturbation equation, with the only effect of a possible modification of the precise numerical values of the parameters $\a_1$, $\a_2$, but {\it not} of their order of magnitude (see for instance \cite{HZ} for a non-perturbative string-background solution with constant curvature). Hence, our subsequent phenomenological analysis of the axion perturbation spectrum may be regarded as implicitly including also all higher-order $\ap$ corrections. 

The primordial axion spectrum amplified by this model of string phase has thus a spectral index determined (according to standard cosmological perturbation theory \cite{15,15a}) by the power-law evolution of the pump field $\xi_\sg$, expressed in conformal time. For the background we are considering we have, in cosmic time,
\beq
\xi_\sg = a \,b^{-n/2} e^{\phi/2} \sim \,e^{\bp t} \,e^{-{n} \gp t/2} \,
e^{{\fp}t/2},
\label{213}
\eeq
for $-\infty <t < + \infty$. Hence, in conformal time where $a \sim (-\bp \eta)^{-1}$,
\beq
\xi_\sg \sim (-\bp\eta)^{-1+ \b_\sg}, ~~~~~~~~~~~~~
\b_\sg = {1\over 2 \dot \b}\left(n \dot \ga-\dot \phi \right) ,
\label{214}
\eeq
where $\eta<0$ for $\bp>0$ and $\eta >0$ for $\bp<0$. 

It should be noted that the canonical equation (\ref{212}) determining the axion spectrum only contains second derivatives with respect to the conformal time, and is invariant with respect to the change of coordinates $\eta \ra -\eta$. The definition of the pump field parameter $\b_\sg$ is thus valid quite independently of the sign of $\bp$, controlling the external-space kinematics (i.e. constant-curvature expansion for $\bp>0$, constant-curvature contraction for $\bp<0$).


\subsection{Tensor perturbations including viscosity}
\label{sec23}

For tensor metric perturbations the procedure is the same as the previous one used for axion perturbations. We start with the action (\ref{21}) and consider the first order perturbation $\da g_{AB}= h_{AB}$ of the metric tensor, expanding the metric as $g_{AB}\ra g_{AB}+ \da g_{AB}$, and computing the perturbed action up to terms of order $h^2$. Assuming as before that the background value of the Kalb-Ramond strength tensor  is vanishing ($H_{ABC}=0$), we shall concentrate on perturbations propagating in the $d=3$ external space, $h_{ij}= h_{ij}(t, x_i)$, and satisfying the usual transverse-traceless (TT) condition: $\pa_j h_i\,^j= 0= h_i\,^i$. Let us first assume that there is no shear viscosity in the matters sources. 
The only contributions to the perturbed action come then from the gravi-dilaton part of the action, and in particular from the terms of Eq. (\ref{21}) not containing the Kalb-Ramond field: the results are known, as already computed in previous papers \cite{19alpha}. 

For the purpose of this paper it will be enough to recall that the quadratic perturbed action for the Fourier components of tensor perturbations, $h_k$, is formally of the same type as in the axion case, Eq. (\ref{28}), and that we obtain exactly the same type of Eqs. (\ref{29}), (\ref{210}) for the propagation of $h_k$ and of its corresponding canonical variable, $u_k=z h_k$.  The only difference is that the effective pump fields $z$ and $y$, for the tensor modes, are different functions of the background variables $a,b,\phi$ (see e.g. \cite{19alpha}). 

However, in the special case of a background with constant values of $\bp, \gp, \fp$ it turns out  again  that the two functions $z$ and $y$ are proportional, and thus define a unique pump field $\xi_h$ for tensor perturbations, in a way similar to Eq. (\ref{211}). One obtains, in particular:
\beq
z \sim y \sim 
\xi_h \sim  a \,b^{n/2} e^{-\phi/2}.
\label{215}
\eeq
The canonical equation, determining the tensor propagation spectrum, then takes the form
\beq
u''_k + \left(k^2c_s^2 - {\xi_h'' \over \xi_h}\right) u_k=0
\label{216}
\eeq
(where again $c_s\sim 1$, but its precise value is in general different from  that of the corresponding parameter of axion perturbations). Let us stress again that, for the particular case of the string-phase background, all higher-order $\ap$ corrections give similar constant contributions to the canonical evolution equations of tensor perturbations, with the only effect of possibly modifying the precise numerical value of the associated phenomenological parameters (like $c_s^2$), but not their order of magnitude (as also stressed before for axion perturbations).

Like in the axion case, we can express the tensor pump field in conformal time, and we obtain:
\beq
\xi_h \sim \,e^{\bp t} \,e^{{n} \gp t/2} \,
e^{-{\fp}t/2} \sim  (-\eta)^{-1+ \b_h}, ~~~~~~~~~~~
\b_h = {1\over 2 \dot \b}\left(\dot \phi -n \dot \ga\right) = -\b_\sg,
\label{217}
\eeq
valid as before for both expanding space, $\bp>0$ with $\eta<0$, and contracting space, $\bp<0$ with $\eta>0$. We note that the obtained relation $\b_h=-\b_\sg$ is an expected consequence of the $S$-duality symmetry \cite{12} satisfied by the model of string phase that we are considering.

The $S$-duality symmetry could be broken, however, if we 
would like to consider a fluid source with non-vanishing shear viscosity. Such an additional``non-minimal" ingredient is not necessarily required, as we shall see, to obtain a scenario compatible with the fit of the PTA data; however, it may be interesting (and useful) to sketch here the basic effects of shear viscosity on tensor perturbations, also in view of the possibility that future detections of relic GW backgrounds, corresponding to unexpected amplitude in unexpected frequency ranges, might be better explained by a spectrum produced in the context of models with broken $S$-duality symmetry.

Indeed, let us recall that the shear viscosity of the sources, unlike bulk viscosity, directly affects the propagation of tensor perturbations, and can thus induce significant  modifications of their spectrum. We refer to \cite{Dolgov} for an introduction to this effect, and for its detailed discussion. Here we only report the modified form of the evolution equation for the Fourier components of the $3$-d tensor perturbation modes $h_k$, in the TT gauge, explicitly written in the same string phase as before (we put for simplicity $c_s =1$), and modified by the presence of a non-vanishing shear viscosity:
\beq
h_k'' + 2 \left({\xi_h'\over \xi_h} + a H_V \right) h_k' +k^2 h_k=0.
\label{218}
\eeq
Here $\xi_h$ is the tensor pump field of Eqs. (\ref{215}), (\ref{217}), and $H_V$ is the parameter depending on the shear viscosity and on the dilaton already defined in Eq. (\ref{26}). 
Note that the explicit form of Eq. (\ref{218}) and, in particular, the presence of the dilaton in the viscosity contribution represented by $H_V$, is due to the direct rescaling of the tensor perturbation equation with viscosity from the Einstein frame used in \cite{Dolgov} to the String frame, used in this paper (see e.g. \cite{15} for the details of the transformations between the two frames, in an arbitrary number of dimensions). It should be stressed, finally, that for the constant-curvature background that we are considering, the parameter $H_V$ is constant (see instead \cite{Fanizza:2022pvx} for an example of time-dependent $H_V$ in more general background geometries).

According to Eq. (\ref{218}), the presence of shear viscosity defines a new effective tensor pump field $\ti \xi_h$ such that 
\beq
{\ti \xi'_h\over \ti \xi_h} = {\xi_h'\over \xi_h} + a\,H_V, 
\label{220}
\eeq
and a new ``viscous" canonical variable $u =\ti \xi_h h_k$. satisfying the generalised equation 
\beq
u_k'' +\left(k^2 - {\ti\xi''_h\over \ti\xi_h}\right)u_k=0.
\label{221}
\eeq
We can  easily find the power-law behaviour of the new, viscous  pump field $\ti \xi_h$ in conformal time from Eq. (\ref{220}). Putting $\ti \xi_h \sim (-\bp \eta)^K$, and solving for $K$, we obtain
\beq
\ti\xi_h(\eta) \sim (-\bp \eta)^{-1+ \ti \b_h}, ~~~~~~~~~~~~~
\ti\b_h = {1\over 2 \dot \b}\left(\dot \phi - n \dot \ga - 2 H_V \right) = 
-\b_\sg - {H_V \over \bp}.
\label{222}
\eeq
As before, the above result for $\ti\xi_h$ is valid for both expanding ($\bp>0, \eta<0$) or contracting ($\bp<0, \eta>0$) external $3$-d space. 

We note, finally that $S$-duality is explicitly broken ($\ti \b_h \not= -\b_\sg$, compare with Eq. (\ref{217})), because shear viscosity affects the tensor pump field and the tensor perturbation spectrum, but not the corresponding parameters of axion perturbations, which are left unchanged. It follows that the phenomenological parameter controlling the $S$-duality violation, defined as $\ep \equiv \b_\sg + \ti \b_h$ in our previous paper \cite{1}, turns out to be directly related to the viscosity of the sources as $\ep = - H_V/\bp$.


\section{Examples of physical models compatible with the PTA signal}
\label{sec3}

Looking at the allowed region of the spectral parameter space presented in \cite{1} it should be noted, first of all, that the conditions to be satisfied by the variables $\bp, \gp, \fp, H_V, \dots$, characterising the chosen background model, are significantly affected by the presence or absence of the $S$-duality symmetry.

Let us recall to this purpose that the overall shape of the today-observed relic GW spectrum is determined not only by the spectral tilt (controlled by $\b_h$) of the modes amplified during the string phase, but also by many other model-dependent details such as the duration of the pre-bouncing evolution, the bouncing energy scale, the durations of the post-bouncing axion-dominated regime and so on (see Appendix \ref{secC}). 
All such physical details will be described in the context of our discussion by phenomenological parameters which, on one hand, control the shape and the amplitude of the spectrum and which, on the other hand, are constrained by theoretical and observational  bounds. Let us only mention here, as an important example, the model-dependent stability of the string-phase solution (discussed, for an isotropic geometry, in \cite{23c}): the associated phenomenological parameter is the time duration of the constant, high-curvature regime preceding the bouncing transition, which corresponds, in its turn, to a related spectral frequency-range for modes crossing the horizon during the string-phase. 

Hence, for any given couple of values of $\b_h, \b_\sg$, their localisation inside or outside the allowed region of a PTA-compatible spectrum also depends on the variation range of all the other parameters (as illustrated in particular by Fig. 1 of ref. \cite{1})). The overall allowed region must of course satisfy,  in addition, all existing phenomenological constraints such as the bounds imposed by Big Bang nucleosynthesis \cite{10}, by the CMB observations \cite{32a}, by the present data of the LKV network \cite{31aa},  by the contribution to the effective number of cosmic relativistic degrees of freedom \cite{ido}, 
 and so on. All such conditions have been summarised in Appendix \ref{secC}, but see also \cite{1} for an explicit and detailed discussion. 

Because of these constraints we have in particular that, when the two pump-field parameters  $\b_h, \b_\sg$ are directly related by $S$-duality, and if we apply on one of them the needed constraints, the other one is also automatically constrained. Conversely, when duality is broken and the two parameters are independent, there is no automatic transmission of constraints, and the allowed range of the two parameters is larger. 
In the first ($S$-dual) case where $\b_h= -\b_\sg$ it turns out, from the results of \cite{1}, that the GW spectral parameter for a scenario compatible with the PTA signal must be confined inside the (rather small) range
\beq
-0.09 \laq \b_h \laq -0.05.
\label{31}
\eeq
In the second ($S$-duality broken) case, like for instance the viscous model of the previous section (with $\ti \b_h \not= -\b_\sg$) one finds instead that the spectral parameters must satisfy two independent conditions:
\beq
-0.08 \laq \ti \b_h \laq 0.05, ~~~~~~~~~~~~~~~
0.10 \laq \b_{\sg} - \ti \b_h \laq 0.21.
\label{32}
\eeq

In spite of these rather strong conditions, we have found various possible examples of fluid sources, with and without dilatonic charge, with and without viscosity, producing a spectrum of tensor and axion perturbations determined by the pump field parameters satisfying the conditions (\ref{31}) or (\ref{32}). In the following subsections we shall present a detailed illustration of simple and physically motivated models of fluid sources compatible with the mentioned constraints, and useful to illustrate the typical properties of the non-minimal scenario. 

Let us notice,  to this purpose, that the above conditions  (\ref{31}),  (\ref{32}) provide constraints on the values of $\bp, \gp, \fp$ (and possibly $H_V$) obtained by solving the background equations (\ref{a1})--(\ref{a4}), and describing a particular example of string phase kinematics. The string phase describes an epoch of pre-bouncing evolution, and thus it is naturally associated to a growing string loop parameter, $\fp>0$. In such a context, it would seem natural to expect also an expanding external space with $\bp>0$, and contracting (or frozen) internal dimensions with $\gp \leq 0$. 

As already mentioned, however, this is not necessarily the case, as explicitly shown by the regular, exact (to all orders in $\ap$) and self-dual models of bounce \cite{cin,Gasperini:2023tus,Conzinu:2023fth}. In that case, indeed, the high curvature regime tends to be characterised  by a sign of the effective Hubble parameter which is exactly the opposite of the sign typical of the asymptotic (initial or final) regimes: namely, a sign which corresponds to a contracting (or even flat) external space, $\bp\leq 0$, and/or to an expanding internal space, $\gp>0$. In the same way, those models also suggests the possibility of a high-curvature pre-bouncing regime with decreasing dilaton $\fp<0$: in such a case, however, the self-consistency of the PBB scenario requires a growth of the effective four-dimensional string coupling $g_4 \sim e^{\phi/2} b^{-n/2}$, namely $\dot g_4 >0$, which implies $\fp > n \gp$.

In the following subsections we will thus consider examples of background solutions satisfying the conditions (\ref{31}) or (\ref{32}) without imposing constraints, a priori, on the sign of  $\bp, \gp, \fp$.


\subsection{Anisotropic fluid sources without viscosity: $S$-dual GW spectrum}
\label{sec31}

The simplest and most natural example of source physically compatible with the high-energy string phase is probably that of radiation, represented by a perfect fluid with traceless stress tensor, i.e. satisfying the condition $3w_1+nw_2 =1$ (see Eqs. (\ref{24d}), (\ref{25})). It might represent the possible effects of the backreaction due to the amplified perturbations, and it could even correspond to an isotropic spatial distribution throughout the entire (external and internal) space, with  equation of state $p= \rho/(3+n)$, namely with $w_1=w_2= 1/(3+n)$. Assuming that there is no viscosity we have no breaking of the $S$-duality for the GW spectrum, and we may look for solutions of the background equations (\ref{a1})--(\ref{a4}) (with $\eta_V=0$) satisfying the stronger condition (\ref{31}), where $\b_h$ is defined by Eq. (\ref{217}). 

In the isotropic case we have checked that there are solution with $\b_h$ varying in the whole allowed range (\ref{31}). However, they need a non-vanishing dilaton charge, and are characterised by numerical values of the parameter $w_3$ which do not seem to have any clear interpretation in terms of physical models of sources.

To this purpose, let us briefly discuss how the dilaton charge density $\sg$  could be physically introduced  for the (possibly anisotropic) fluid model of source that we are considering. First of all we note that the the canonical stress tensor (\ref{24b}), without viscosity, can be easily derived from the following effective action, 
\beq
S_m= -{1\over 2} \int d^{4+n}x \sqrt{-g} \left[ \left((\r +p\right) g_{AB}\,u^A u^B +\left(q-p\right) g_{AB}\,v^A v^B- (\r+3p) - n (p-q) \right],
\label{33}
\eeq
which generalises to our higher-dimensional metric  (\ref{22}) the action for an isotropic fluid given in \cite{Ford}. 
By applying the canonical definition (\ref{24a}) and specifying, after the variation, the vectors $u^A$ and $v^A$ in the comoving gauge, we obtain indeed the previous results (\ref{24d}), (\ref{25}) with $\eta_V=0$.

Suppose now we introduce for this model of fluid the coupling to the dilaton, according to the standard string coupling expansion: namely by multiplying the Lagrangian of the action (\ref{33}) by $e^{k\phi}$, where $k=-1$ for the small coupling limit of the tree-level contribution, $k=0$ for the one-loop contribution, and so on. We then obtain a dilaton-dependent action $S_m(\phi)$ with the same form of Eq. (\ref{33}), but with rescaled fluid variables: 
\beq
\r \ra \r (\phi) = e^{k\phi} \r, ~~~~~~~~~~~
p \ra p (\phi) = e^{k\phi} p, ~~~~~~~~~~~
q \ra q(\phi) = e^{k\phi} q.
\label{35}
\eeq
In that case, the canonical definition of the dilaton charge (see Eq. (\ref{24a})) immediately gives
\beq
{\sg\over 2} =- k p = -k w_1 \r \equiv w_3 \r.
\label{36}
\eeq
At the tree-level, in particular, we have $w_3=w_1$. We shall now consider three simple examples of sources.

\noi
\textbullet \,  {\bf Radiation-like sources}

\noi
Coming back to our radiation-like model of source, and considering the anisotropic case which satisfies the condition $3 w_1+nw_2=1$, we are now physically motivated to look for solutions with no dilaton charge, $w_3=0$, or with tree-level charge, $w_3=w_1$. In both cases we find that there are non-trivial solutions to Eqs.  (\ref{a1})--(\ref{a4}), for the four variables $C,\bp, \gp,\fp$, and with $C>0$, for any given value of $\b_h$ in the range (\ref{31}). 

If we assume $w_3=0$ we find that the solutions are characterised by opposite values of the sign of $\bp$ and $\gp$. However, if we want to obtain a growing dilaton and $4$-dimensional string coupling, $\fp>0$, $\dot g_4 >0$, then we are left with the case $\bp<0$ and $\gp >0$, namely (as anticipated) with a solution describing contracting external space and expanding internal space, just in agreement with the high-curvature kinematics of the previously mentioned  regular bouncing scenarios  \cite{cin,Gasperini:2023tus,Conzinu:2023fth}. We have checked that the above properties hold for arbitrary numbers $n$ of internal dimensions. If we choose, for instance, the typical superstring value $n=6$, and the particular (allowed) value $\b_h= -0.06$ for the spectral parameter, we obtain
\beq
\b_h=-0.06,~~~~~~C \simeq 2.20, ~~~~~~\bp \simeq -0.063, ~~~~~~ \gp \simeq 0.30, ~~~~~~ \fp \simeq 1.80, ~~~~~~
\dot g_4 >0.
\label{37}
\eeq

If we assume instead $w_3=w_1$ (tree-level dilaton charge), we find for the allowed solutions that $\bp$ and $\gp$ must have the same sign, while $\fp$ and $\dot g_4$  the opposite one: and if we choose, in particular, a growing dilaton, $\fp>0$, we obtain that the external and internal space must be expanding, $\bp>0, \gp>0$, while $\dot g_4<0$.  The opposite is true if we choose $\fp<0$. For the typical value $n=6$, and, again, for $\b_h = -0.06$, we obtain, for instance:
\beq
\b_h=-0.06,~~~~~~C \simeq 0.020, ~~~~~~\bp \simeq 0.037, ~~~~~~ \gp \simeq 0.23, ~~~~~~ \fp \simeq 1.39, ~~~~~~
\dot g_4 <0.
\label{38}
\eeq

\noi
\textbullet \,  {\bf Unstable strings}

\noi
Another possible example of source, also typical of the high-curvature string phase, may correspond to the presence of a stochastic distribution of primordial unstable strings \cite{Gasperini:1990xg, Gasperini:1991rv}, described by a gas which in the isotropic $D$-dimensional case is characterised by the averaged equation of state $p=-\r/(D-1)$, i.e. by $w_1=w_2=-1/(3+n)$. In such a case we can find background solutions compatible with the condition (\ref{31}) but, like in the case of isotropic radiation, the needed value of the dilaton charge seems to have no direct physical interpretation. In particular, there are no solutions with a value of $w_3$ which is nonvanishing and compatible with the model of coupling described by Eq. (\ref{36}). 

However, for an anisotropic distribution of unstable strings (without viscosity) whose stress tensor, of the type (\ref{25}), satisfies the generalised condition $3w_1+nw_2=-1$, the results are different. There are indeed background solutions compatible with the range of Eq. (\ref{31}), and sourced by anisotropic unstable strings with tree-level dilaton charge: namely by a fluid with $w_2= -(1+3w_1)/n$ and $w_3=w_1$. The properties of such solutions are very similar to those of the previously obtained solutions sourced by a charged radiation fluid: indeed, we have again   $\bp>0$, $\gp>0$, $\fp>0$, $\dot g_4<0$ (like in Eq. (\ref{38})). Interestingly enough, we recall that the presence of a fully expanding (external and internal) space, in this case, is not an optional (or casual) property of the obtained geometry, but an unavoidable result needed for the consistency of the unstable-string model \cite{Gasperini:1990xg, Gasperini:1991rv}. To give here an example of explicit solution we can choose, as before, $n=6$, $\b_h=-0.06$, and we find
\beq
\b_h=-0.06,~~~~~~C \simeq 0.017, ~~~~~~\bp \simeq 0.036, ~~~~~~ \gp \simeq 0.23, ~~~~~~ \fp \simeq 1.37, ~~~~~~
\dot g_4 <0.
\label{39}
\eeq

\noi
\textbullet \,  {\bf String-hole gas}

\noi
Let us finally consider a third example of fluid source, which is probably the most natural and physically motivated one possibly present in the string phase: the so-called string holes gas (SHG), namely a spatially isotropic distribution of high-energy, string-size black holes, typically formed when the curvature reaches the string scale (see eg. \cite{23a,23b,23c}). The isotropy gives $p=q$, and the string hole condition implies, in the String frame,  $p= \sg/2$. Hence $w_1=w_2=w_3$.

 We can then consider the particular case $w_i=-1$, which also corresponds to a very simple (and possibly interesting) physical interpretation of the source in terms of an exponential dilaton potential. 
Indeed, let us start with the following matter action $S_m$,
\beq
S_m = - \int d^D x \sqrt{-g} \,V_0 e^{-\phi}, ~~~~~~~~~~~~~~~V_0= {\rm const},
\label{310}
\eeq
and apply the previous canonical definitions of the stress tensor and of the  dilatonic charge. We immediately obtain 
\beq
T_A\,^B=  \da_A^B\,V_0 \,e^{-\phi}, ~~~~~~~~~~~~~~~~
{\sg\over 2}= -  V_0 \,e^{-\phi},
\label{311}
\eeq
which implies, using the fluid model of Eq. (\ref{25}),  $w_1=w_2=w_3=-1$.

Quite independently of its possible interpretation, even in this case we obtain non-trivial solutions of Eqs. (\ref{a1})--(\ref{a4}), with $\eta_V=0$,  for any value of $\b_h$ compatible with Eq. (\ref{31}), and for arbitrary values of $n$. Again  we find that a growing dilaton, $\fp>0$, must be associated to  an expanding (internal and external) geometry, $\bp>0, \gp>0$, but, in this case, even the four-dimension string coupling turns out to be growing, $\dot g_4>0$. For the case $n=6$ and $\b_h =-0.06$ we obtain, for instance, 
\beq
\b_h=-0.06,~~~~~~C \simeq 0.009, ~~~~~~\bp \simeq 0.035, ~~~~~~ \gp \simeq 0.22, ~~~~~~ \fp \simeq 1.33, ~~~~~~
\dot g_4 >0.
\label{312}
\eeq

\begin{figure}[t]
\centering
\includegraphics[width=14cm]{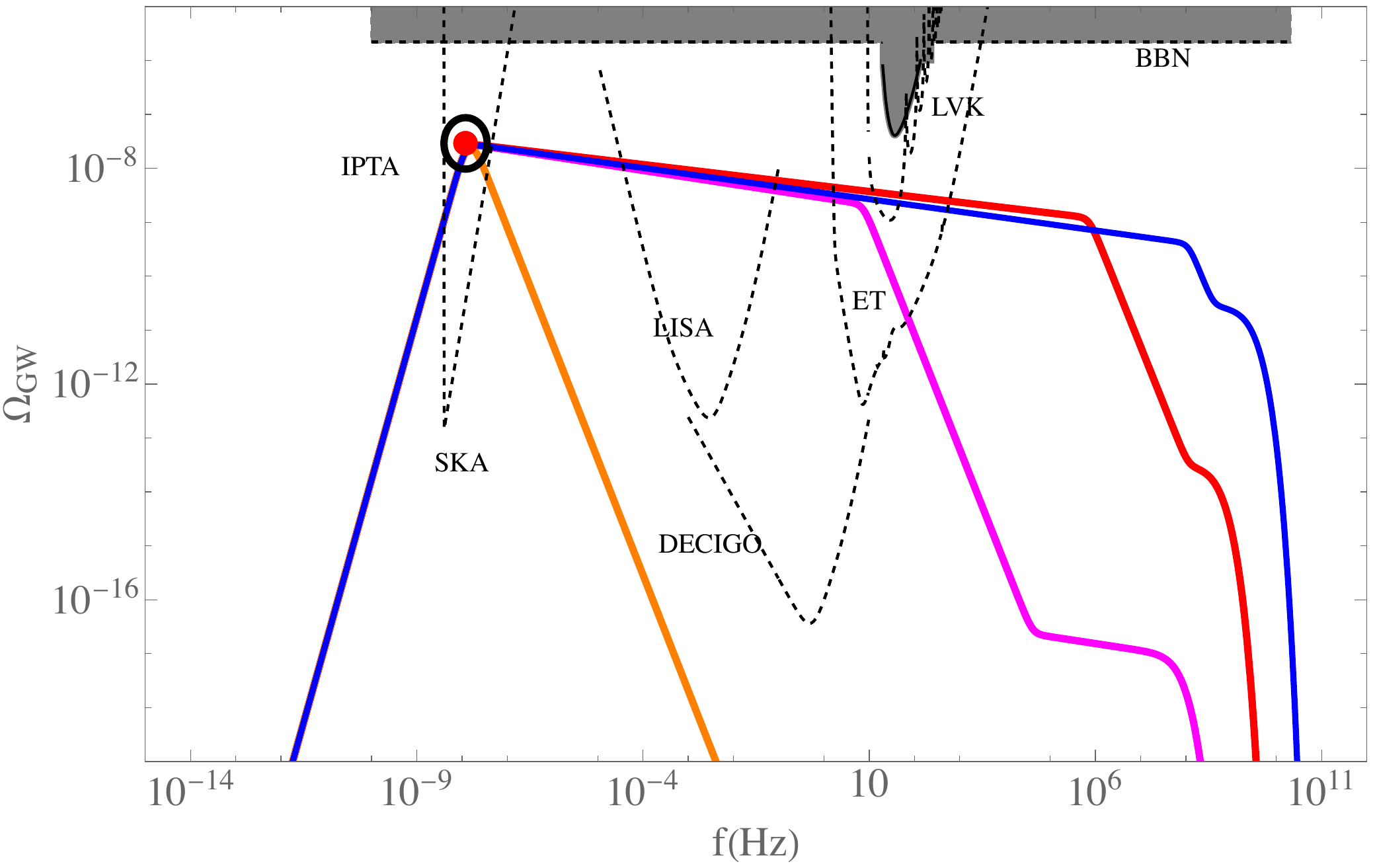}
\caption{Examples of GW spectra produced by non-minimal but still $S$-dual models, with a string phase described by the background solutions of Sect. \ref{sec31}. All plotted spectra  satisfy the constraints of Appendix \ref{secC}, but with different values of $\b_h$ and $\sg_i/\Mp$. In particular: the red spectrum corresponds to $\b_h=-0.05$ and  $\sg_i/\Mp=1$; the blue spectrum corresponds to $\b_h=-0.058$ and  $\log (\sg_i/\Mp)=-1/2$; the orange spectrum corresponds to $\b_h=-0.073$ and  $\log(\sg_i/\Mp)=-1/2$. 
 Also shown is a typical $S$-dual spectrum with $\b_h=-0.064$ and  $\log(\sg_i/\Mp)=-1/2$ (the magenta plotted curve), providing an explicit example of how this class of models, beside explaining the already detected signal, could also produce signals in the sensitivity range of future detectors such as LISA, ET and DECIGO.
}
\label{f1}
\end{figure}
All the previous examples of backgrounds, sourced by isotropic or anisotropic high-energy fluids, with or without dilaton charge, and without viscosity, are able to produce a GW spectrum (and an associated, $S-$duality-related axion spectrum) which, in the frequency band amplified by the string phase, has a spectral power controlled by the parameter $\b_h$ satisfying the condition (\ref{31}). 
See Appendix \ref{secC} for the explicit form of the spectral energy density $\Om_{GW}(f)$ written in terms of $\b_h$ and of the other parameters. 
Of course, as previously mentioned, the today observed GW spectrum, and in particular a spectrum responsible of the detected PTA signal, must satisfy many other phenomenological and 
model-dependent constraints reported in Appendix \ref{secC} and discussed in details in \cite{1}. 
We have shown in Fig. \ref{f1} a few example of such spectra compatible with all the imposed constraints, and chosen in such a way as to qualitatively represent the possibly allowed  larger amplitude (the red plotted curve), the possibly allowed  higher frequency extension (the blue plotted curve), and the limiting case in which the string phase ends with a  bouncing transition just after the horizon-exit of the modes producing the PTA signal (the orange plotted curve).

In that figure $\Om_{GW} = f d \rho_{GW}/ (\rho_c df)$ is the spectral energy density in units of critical energy density $\rho_c$, $f$ is the today observed proper frequency of the wave modes, and the IPTA signal (with $\Om_{GW} \simeq 2.9 \times 10^{-8}$, $ f \simeq  1.2 \times 10^{-8} {\rm Hz}$) is denoted by the red dot localised inside the black circle. The low-energy spectrum amplified before the beginning of the string phase ($f<f_s$) has a steep power-law growth, $\Om_{GW} \sim f^3$, represented by the steep solid line (common to all the plotted spectra) reaching the red dot at $f=f_s$. 

We have also inserted into the figure the expected sensitivities of near-future GW detectors like SKA \cite{ska}, LISA \cite{27}, ET \cite{28}, DECIGO \cite{29} (corresponding to the regions of the spectral plane inside the black dashed curves). It may be interesting to note that the spectra predicted by these examples of non-minimal models are possibly well inside the expected experimental sensitivities. 

Finally, the grey-shaded area describes the upper bounds  presently imposed by the LKV network \cite{31aa} and by Big Bang  nucleosynthesis (BBN) \cite{10},  always automatically satisfied. It may be noted that also the bound  concerning the GW contribution to the cosmic number of relativistic degrees of freedom, 
$\Da N_{\rm eff}$, proportional to the GW spectral energy density (in critical units) integrated over the full frequency range (see eg. \cite{ido})  
-- and thus closely related to the BBN bound -- 
turns  out to be easily satisfied by the class of models we have considered.


\subsection{Viscosity and broken $S$-duality}
\label{sec32}

To give an example of non-minimal pre-big bang scenario with broken $S$-duality symmetry we shall now assume that  the fluid source contributing to the dynamics of the high-curvature string phase has a non-vanishing shear viscosity, $\eta_V \not=0$, as explicitly taken into account in Sect. \ref{sec21} (see in particular Eqs. (\ref{24b}), (\ref{24d})). 
 As already stressed in Sec. \ref{sec21}, we expect such viscosity to produce a corresponding spectral parameter $H_V$ of the order of the string mass scale. Indeed, shear viscosity may typically appears at high curvature, as recently discussed, in particular, for the SHG case \cite{Quin} (but the presence of viscosity can affect, of course, also other models of sources). In such a case the slope of the high-frequency GW spectrum, controlled by $\ti \b_h$,  would be modified as discussed in Sect. \ref{sec23}, while the slope of the associated axion spectrum, controlled by $\b_\sg$, would keep unchanged: one would then obtain $\ti \b_h \not=-\b_\sg$ and $\ep = \b_\sg+ \ti \b_h \not=0$, according to Eq. (\ref{222}). We have, in particular, two possibilities.  

The first one is a string phase with an isotropic geometry, $\bp= \gp$. In that case the shear viscosity does not contribute to the background solution \cite{Dolgov} (see Eq. (\ref{24d})) but only to the evolution of tensor perturbations. Hence, the shear parameter $H_V$ is in principle free, and can be modeled {\it ad hoc} to try to satisfy the two conditions (\ref{32}), once the background values of $\bp=\gp$, $\fp$ and $C$ are given.

The second one refers, as before, to an anisotropic geometry, $\bp \not=\gp$. In that case the viscosity explicitly appears in the stress tensor  (see again Eq. (\ref{24d})), hence it also contributes to the background equations of motion, and it is no longer a free parameter but it turns out to be fixed by the given chosen solution. 

Considering the second, more general possibility we have explicitly checked that allowed $S$-duality breaking solutions, giving a non-minimal spectrum compatible with all constraints, can be obtained by adding the shear-viscosity contribution to any one of the three types of fluid sources considered in the previous section (radiation, unstable strings, SHG). However, the corresponding allowed values of  $\b_\sg$, $\ti \b_h$ do not necessarily  cover the whole range of values defined by Eqs. (\ref{32}) for all types of sources.

For brevity, we present here a single, straightforward example involving a fluid source that describes anisotropic radiation with shear viscosity and no dilaton charge: a fluid source with equation of state $3w_1+nw_2=1$, and $w_3=0$. We choose this model, in particular, to emphasise the main possible difference with the $S-$dual examples of the previous section: namely, a GW spectrum with a blue (i.e. growing) spectral behaviour in the string phase, which requires $\ti \b_h >0$ (see the spectral distribution (\ref{c1})).
Indeed, for the other two previous examples (string holes and unstable strings) there are no allowed background solutions with this property, even assuming that their viscosity is non-vanishing.

Assuming as before $n=6$ we can take, for instance, $\ti \b_h=0.01$, $\b_\sg - \ti \b_h = 0.14$, just to use values located around the middle of the allowed region (\ref{32}), and we then obtain
\beq
\ti \b_h=0.01,~~~~~\b_\sg - \ti \b_h = 0.14, 
~~~~~C \simeq 2.20, ~~~~~H_V \simeq 0.009, 
~~~~~\bp \simeq -0.06, ~~~~~ \gp \simeq 0.30, ~~~~~ \fp \simeq 1.81.  \label{314}
\eeq
It should be noted that for such solution (but also for solutions corresponding to different values of the spectral parameters) the dilaton is growing (as expected),  and, just like in the previous examples without viscosity, the solutions again describe a string phase with contracting external space and expanding internal space ($\bp<0$, $\gp >0$), just as predicted by the  high-curvature kinematics of the regular bouncing scenarios \cite{cin,Gasperini:2023tus,Conzinu:2023fth}.

\begin{figure}[h]
\centering
\includegraphics[width=14cm]{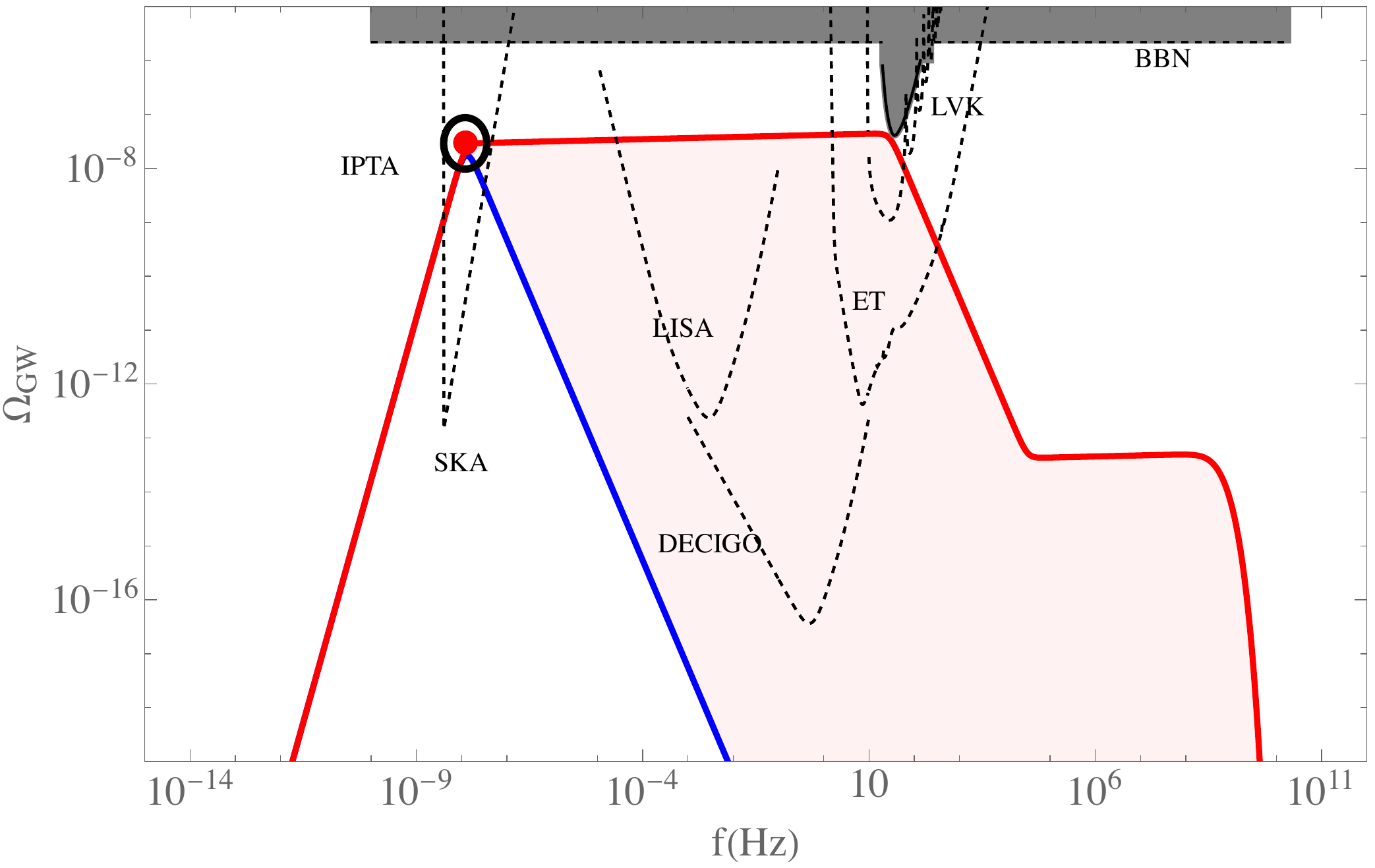}
\caption{The allowed region (the pink shaded area) spanned by the GW spectra  produced by a string phase described by the solution (\ref{314}). The two limiting GW spectra, represented by the red and  blue curves, are both violating the standard $S-$duality symmetry with the same symmetry breaking parameter, $\ep=0.16$. However, the red curve corresponds to $\log(\sg_i/\Mp)=-0.81$, the blue one corresponds to $\log(\sg_i/\Mp)=-0.49$.}
\label{f2}
\end{figure}

Given the above values of the spectral parameters, and taking into account all spectral constraints reported in Appendix \ref{secC}, we are only left  (as in the previous duality-invariant case) with the freedom of shifting the value of the axion-parameter $\sg_i/\Mp$ using, for instance, the allowed variation range assumed in \cite{1}. 
We obtain in this way a corresponding allowed region for the GW spectrum of this non-minimal, non-dual model illustrated by the pink shaded area of Fig. \ref{f2}, and limited by the spectrum with minimal (the blue curve) and maximal (the red curve) extension of a string phase producing a slightly growing spectrum with spectral parameter $\ti \b_h=0.01$.

It should be stressed, finally, that the plotted spectra and the corresponding allowed region are only a very 
simple example of a non-dual and non-minimal model based on high-curvature background equations modified by the presence of viscous, anisotropic, radiation-like sources. 
By no means they are to be regarded as describing the general predictions of a non-minimal scenario with duality-broken symmetry.
Such a general discussion is outside the object of this short paper, and will be possibly discussed in future works.


\section{Conclusion}
\label{sec4}

In this paper we have presented, for the first time (we believe) in the literature of theoretical cosmology, two new results: $i)$ the explicit computation of the  perturbation equation for the Kalb-Ramond axion field including 
higher-curvature corrections to first order in $\ap$; and $ii)$ the explicit contribution of shear viscosity to the tensor metric perturbations of the gravi-dilaton equations written in the String frame. 

We have used such results to discuss examples of high-curvature, non-vacuum, non-minimal pre-big bang models giving a spectral distribution of relic stochastic gravitons, $\Om_{GW}(f)$, which may be the source of the signal recently detected by the IPTA collaboration. We have shown, also, that the GW distribution of the considered models, in the spectral plane $\{f, \Om_{GW}(f)\}$, may have an intensity   
possibly detectable even in the higher-frequency sensitivity window of the so-called third generation GW antennas, such as LISA, ET and DECIGO.  It would be interesting to perform independent fits of the PTA data and compare the resulting spectrum with the class of model-dependent spectra discussed in this paper, to confirm (or disprove) the identification of the string frequency scale $f_s$ with the typical frequency scale  of the PTA signal (see Eq. (\ref{c4})).

Finally, we have presented a mechanism able to break the string theory $S$-duality relating the primordial graviton and axion spectra. The breaking may be induced by the intrinsic presence of shear viscosity in the fluid sources contributing to the dynamics of the high-curvature string phase, and it could be  important to interpret (and explain) possible future signals of relic GW backgrounds detected with unexpected amplitude in atypical (high-energy) frequency ranges.

\section*{Acknowledgements} 

It is a pleasure to thank Giuseppe Fanizza, Luigi Tedesco and Gabriele Veneziano for earlier collaborations on the general aspects of the non-minimal pre-big bang scenario and of the allowed spectrum (including the possible viscosity contributions). 
We are grateful to Gianluca Calcagni for providing us with the sensitivity data of the gravitational antennas illustrated in the figures of this paper. PC is supported in part by INFN under the program InDark: {\it ``Inflation, Dark
Matter and the Large-Scale Structure of the Universe"}.
MG and EP are supported in part by INFN under the program TAsP: {\it ``Theoretical Astroparticle Physics"}. EP is  also supported by the research grant number 2022E2J4RK {\it ``PANTHEON: Perspectives in Astroparticle and Neutrino THEory with Old and New messengers"}, under the program PRIN 2022 funded by the Italian Ministero dell'Universit\`a e della Ricerca (MUR) and by the European UnionNext Generation EU.

%

\appendix
\section{Field equations for the non-minimal, non-vacuum string phase}
\label{secA}

Here we give explicitly the equations controlling the time evolution of the string phase background, describing the chosen model of non-minimal PBB scenario. 

Let us start with the action (\ref{23}), and with the sources described by Eqs. (\ref{24a})--(\ref{25}). We vary the action with respect to $N, \b, \ga, \phi$ and, after the variation, we put $N=1$ (cosmic time gauge), and $\ddot \b=0$, $\ddot \ga=0$, $\ddot \phi =0$ (to obtain a constant spacetime curvature and a linearly evolving dilaton, as in the minimal case). We then obtain, respectively, the following  four algebraic equations:
\bea
&&
\fp^2+ 6 \bp^2 +n(n-1)\gp^2+ 6 n \bp \gp - 6 \bp\fp-2n \gp \fp-
\nonumber \\&&
- 3 \ap a_0 \left[ c_2 \gp^4+ 8\fp\bp^3 + c_4 \gp^3(\fp-3 \bp)- \fp^4 + c_5\bp\gp^2(\fp-\bp)+8 n \gp \bp^2(3 \fp- \bp)\right]=
\nonumber \\&&
= 2 \ls^{D-1} e^\phi \rho,
\label{a1}
\eea
\bea
&&
-\fp^2- 6 \bp^2  +4 \bp\fp  +2 n \gp(\fp-2\bp)-n(n+1)\gp^2
\nonumber \\&&
- \ap a_0\left[8 \fp^2\bp^2 - 16 \fp \bp^3 - \fp^4\right]- \ap a_0 \gp^4(c_2+nc_4)
+ {n\over 3} c_5 \ap a_0 \gp^3 (\fp- 2 \bp)
\nonumber \\&&
+ \ap a_0 \gp^2 \left[- \bp^2(c_5+8 n^2 ) +\bp \fp\left({2\over 3} c_5+16 n^2  \right) -{c_5\over 3} \fp^2 \right]
+8n \ap a_0 \gp\left[5 \bp^2 \fp-2 \bp^3-2 \bp \fp^2 \right]=
\nonumber \\&&
= 2  \ls^{D-1} e^\phi \left[ w_1 \rho +{2 \eta_V\over 3+n}{n\over \ls} (\gp-\bp)\right],
\label{a2}
\eea
\bea
&&
 -n^2(n-1) \gp^2 -6 n (n-1) \gp \bp + 2n(n-1)\gp \fp +6 n\bp\fp-n\fp^2-12n \bp^2
\nonumber \\&&
+ \ap a_0 3 n c_2 \gp^4 
+ \ap a_0 2\gp^3(\fp- 3 \bp)(n c_4 - 2 c_2)
+\ap a_0\gp^2\left[nc_5\bp(\fp-\bp)-3 c_4(\fp-3\bp)^2 \right]
\nonumber \\&&
+\ap a_0\gp\left[c_5(\fp-\bp)(6\bp^2-2\fp\bp)\right]
-\ap a_0 n \left[-72\fp\bp^3-\fp^4+24\bp^2(\fp^2+\bp^2)\right]=
\nonumber \\&&
= 2 n   \ls^{D-1} e^\phi \left[ w_2 \rho +{2 \eta_V\over 3+n}{3\over \ls} (\bp-\gp)\right],
\label{a3}
\eea
\bea
&&
\fp^2+12\bp^2-6\bp\fp+n(n+1)\gp^2+2n\gp(3\bp-\fp)
+\ap a_0\left[24\bp^4+3 \fp^4-12\bp\fp^3\right]
\nonumber \\&&
+ \ap a_0\left[ \gp^4(c_2+nc_4)+ \gp^3n c_5 \bp+\gp^2\bp^2(2 c_5+24n^2 )+n \gp(72\bp^3-4\fp^3)\right]=
\nonumber \\&&
=- 2w_3 \ls^{D-1} e^\phi \rho.
\label{a4}
\eea
For all the applications of this paper we shall use the convention $\ap a_0=1/4$ (bosonic string model in units $\ap=1$ and heterotic superstrings in units $\ap=2$), and we shall define the constants $C$ and $H_V$, related to the energy density and viscosity of the matter sources as well as to the dilaton, as  specified in Eq. (\ref{26}).

\section{Axion perturbations in the string phase background}
\label{secB}

Let us start with the $\ap$-corrected action (\ref{21}), and perturb the NS-NS two-form $B$ by putting $H_{ABC} \ra H_{ABC}+ \da H_{ABC}$, where $\da H$ is defined in terms of the Kalb-Ramond axion $\chi$ according to Eq. (\ref{27}). Let us assume that $\chi= \chi(t,x_i)$, that the zeroth-order Kalb-Ramond background is vanishing ($H_{ABC}=0$), and that the matter action $S_m$ is decoupled from $B$. Working with the explicit form (\ref{22}) of the metric tensor, but using the convenient conformal time coordinate $\eta$ defined by $dt= a d\eta$, we then obtain the following quadratic perturbed action 
\beq
\da_H S = -{1\over 2 \ls^2} \int d \eta \, a^2 b^{-n} e^{\phi} \left[ A(\eta) \pa_\mu \chi \pa^\mu \chi +B^{\mu\nu}(\eta) \pa_\mu \chi \pa_\nu \chi \right],
\label{b1}
\eeq
where
\bea
&&
A= {1\over 2} \left[ 1 + \ap \a_0 \left(g^{MN} R_{AMN}\,^A + 4 g^{\mu\nu} R_{A\mu\nu}\,^A +2 g^{\mu\nu} R_{\a\mu\nu}\,^\a+11 \nabla \phi^2 \right)\right],
\nonumber \\ &&
B_{\mu\nu} = 2 \ap a_0 \left( \pa_\mu \phi \pa_\nu \phi - R_{A\mu\nu}\,^A - 
 R_{\a\mu\nu}\,^\a \right). 
\label{b2}
\eea

Note that in the above equations we have explicitly written the coefficients $\ap a_0$ for a clear identification of the high-curvature $\ap$ contributions. We also recall that capital Latin indices run from $0$ to $D-1$, Greek indices from $0$ to $3$. Finally, it should be noted that the three Ricci tensors $R_{AMN}\,^A$, $R_{A\mu\nu}\,^A$ and $R_{\a\mu\nu}\,^\a$, appearing in Eq. (\ref{b2}), are in general all different, as they contain different contributions of the internal and external geometry. We have, for our metric (\ref{22}):
\bea
&&
R_{\a 00}\,^\a=-3(\ddot \b + \bp^2),  ~~~~~~~~~
R_{\a ij}\,^\a = - g_{ij} (\ddot \b + 3\bp^2), ~~~~~~~~
R_{A00}\,^A= -3(\ddot \b + \bp^2)- n (\ddot \ga +\gp^2),
\nonumber \\ &&
R_{A ij}\,^A= -g_{ij}(\ddot \b + 3\bp^2+ n \gp \bp),
~~~~~~~~~~~
R_{Amn}\,^A=- g_{mn}(\ddot \ga + n \gp^2 +3 \bp \gp). 
\label{b3}
\eea

For a more explicit form of the action (\ref{b1}) controlling the axion dynamics we need, in particular, the functions $A, B_{00}, B_{ij}$. By using Eqs. (\ref{b2}), (\ref{b3}) we find that the effective action for the Fourier components $\chi_k$ of the axion perturbations can be put in the form of Eq. (\ref{28}) with
\beq
z^2(\eta)= a^2 b^{-n} e^{\phi} f_1(\eta), ~~~~~~~~~~~~~~~
y^2(\eta)= a^2 b^{-n} e^{\phi} f_2(\eta), 
\label{b4}
\eeq
where
\bea
&&
f_1(\eta)= {1\over 2} \left[  {15\over 2} \fp^2 - 9 \ddot \b - n \ddot \ga 
-30 \bp^2-{n\over 2} (n+1) \gp^2- 9 n \gp \bp 
\right],
\nonumber \\ &&
f_2(\eta)= {1\over 2} \left[   {11\over 2} \fp^2 - 17 \ddot \b - 3n \ddot \ga 
-30 \bp^2-{n\over 2} (n+5) \gp^2- 7 n \gp \bp 
\right].
\label{b5}
\eea
Finally, for the particular background we are considering, we have to impose $\ddot \b=0= \ddot \ga$. In this limit $f_1$ and $f_2$ are constant ($f_1= \a_1^2$, $f_2= \a_2^2$), the two pump fields $z$ and $y$ become proportional as anticipated in Eqs. (\ref{211}), and the final canonical equation for the axion perturbations takes the form (\ref{212}), with
\beq
c_s^2 = \left(\a_1\over \a_2\right)^2=1+{2\over \a_2^2}\left( \fp^2+n \gp^2-2 n \gp \bp \right).
\label{b6}
\eeq

\section{The non-minimal relic GW spectrum and the associated constraints}
\label{secC}

We briefly report here, for the reader's convenience, the explicit form of the relic GW spectrum obtained in the context of a non-minimal pre-big bang scenario, and the phenomenological and theoretical constraints to be imposed for the consistent interpretation of such a spectrum as a possible source of the observed PTA signal. 

Let us first recall, for a short but more complete introduction to the details of our spectral computation, that a successful matching of the various spectral branches (including the high-curvature bouncing transition from pre-to post big bang) has been obtained by applying the so-called ``sudden  approximation" \cite{Gar}, and imposing on the pump field to be continuous at the various transition scales. This has been possible because, in the considered scenario, the transition epochs from one cosmic phase to another are of very short, negligible duration 
compared with the time extension of such phases (including the case of the high-curvature bouncing transition, see e.g.  \cite{GGV1,GGV2}). This produces, however, a spectral distribution of  broken power-law type, and for a better graphical illustration of such spectra (see the plots of Fig. \ref{f1}, Fig. \ref{f2})  we have also applied  a smooth interpolation of all the branches by using the formalism discussed in \cite{7}. 
By applying the above procedure we have obtained the followiwing spectrum. 

The today observable energy density of the spectrum $\Om_{GW}(f,t_0)= \rho_c(t_0)^{-1} d\rho_{GW}(t_0)/(d \ln f)$, given in units of critical energy density $\r_c(t_0)$, has in general four high-frequency branches corresponding to modes living the horizon in the initial dilaton-driven phase or in the string phase, and re-entering the horizon in the dust-like phase dominated by the oscillations of the axion background or in the standard radiation era. It can be parametrised as follows \cite{1,18,7}:
\beq
	\!\!\!\!\!\!\!\!\!\!\!\!\!\!\! \!\!\!\!\!
	\frac{\Om_{\textsc{gw}}(f,t_0)}{\Om_{\textsc{gw}}(f_1,t_0)}  = 
	\begin{cases}
        \left(\dfrac{f}{f_1}\right)^{3- |3-2 \b_h|}, &f_\sg \laq f \laq f_1 \\ \\
		\left(\dfrac{f_\sg}{f_1}\right)^{3- |3-2 \b_h |}
		\left(\dfrac{f}{f_\sg}\right)^{1- |3-2 \b_h|},\qquad  &f_d \laq f \laq f_\sg \\ \\
		 \left(\dfrac{f_\sg}{f_1}\right)^{3- |3-2 \b_h |} \left(\dfrac{f_d}{f_\sg}\right)^{1- |3-2 \b_h |}
		\left(\dfrac{f}{f_d}\right)^{3- |3-2 \b_h |},\qquad
        &f_s \laq f \laq f_d\\ \\
 \left(\dfrac{f_\sg}{f_1}\right)^{3- |3-2 \b_h |}\left(\dfrac{f_d}{f_\sg}\right)^{1- |3-2 \b_h |}
\left(\dfrac{f_s}{f_d}\right)^{3- |3-2 \b_h |}
		\left(\dfrac{f}{f_s}\right)^{3},\qquad \,\,\,& f \laq f_s\,\,.
	\end{cases}
	\label{c1}
\eeq
Here
\beq
\Om_{\textsc{gw}}(f_1,t_0)= \Om_r(t_0) \left(H_1\over \Mp\right)^2 \left(f_d\over f_\sg\right)^2,
\label{c2}
\eeq
where $ \Om_r(t_0)\sim 10^{-4}$ is the present critical fraction of radiation energy density (including neutrinos), and $H_1$ (in units of Planck mass $\Mp$) is the curvature ``bouncing scale" marking the transition to the decelerated post-inflationary evolution. The spectral parameter $\b_h$
is the same as given in Eq. (\ref{217}); obviously, in the presence of viscosity ($H_V \not=0$), it has to be replaced by the generalised parameter $\ti \b_h$ of Eq. (\ref{222}). 
Finally, $f_s, f_1, f_\sg, f_d$ are transition frequency scales marking, respectively, the beginning and the end of the string phase, and the beginning and the end of the post-bouncing, axion-dominated regime. They are related to the physical parameters represented the bouncing scale $H_1$, the initial value $\sg_i$ of the post-bouncing axion background, and the mass $m$ of the oscillating axion, and can be conveniently expressed in terms of the dimensionless parameters $z_s,z_d, z_\sg$ as follows:
\beq
z_s ={f_1\over f_s}, ~~~~~~~
z_d ={f_1\over f_d}, ~~~~~~~
z_\sg ={f_1\over f_\sg}, ~~~~~~~~~~~~~
z_s \gaq z_d > z_\sg \gaq 1.
\label{c3}
\eeq

Let us now give the full list of constraints to be imposed on the spectral distribution (\ref{c1}) and on its parameters. 

First of all, in order to reproduce the PTA signal, we shall assume that \cite{cin1}
\beq
\Om_{\textsc{gw}} (f_{s},t_0) \simeq 2.9 \times 10^{-8}, ~~~~~~~~~~~~~~~~~
f_{s} \simeq 1.2 \times 10^{-8} {\rm Hz}.
\label{c4}
\eeq
Since the (non-dual) spectrum may be growing in the string phase,  $f>f_s$, we have thus to take into account also the experimental upper bound on the spectrum imposed by recent data of the LIGO-Virgo-KAGRA (LVK) network \cite{31aa}, namely:
\beq
\Om_{\textsc{gw}}(f_{\rm LVK})< 4.12 \times 10^{-8} ,~~~~~~~~~~~~
f_{\rm LVK} \simeq 35.4\,{\rm Hz}\,. 
\label{c5}
\eeq

From the theoretical side, on the other hand, we have constraints on the  values of the physical parameters $H_1, \sg_i, m$ which control, respectively, the high-curvature bouncing scale, the scale $H_\sg \sim m (\sg_i/Mp)^4$ of the dust-like axion-dominated oscillations, and the axion-decay scale $H_d \sim (m/\Mp)^2$. Here  
we adopt, in particular, the same constraints used in \cite{1}:
\beq
10^{-3} \laq {H_1\over \Mp} \laq 10^{-1}, ~~~~~~~~~~~~~
10^{-3/2} \laq {\sg_i\over \Mp} \laq 1,
\label{c6}
\eeq
which, by expressing $\sg_i$ and $m$ in terms of the parameters $z_i$ of Eq.  (\ref{c3}),
\beq
{m\over \Mp} \simeq \left(H_1\over \Mp\right)^{1/3} z_d^{-1} z_\sg^{1/3},
~~~~~~~~~~~~~~~~~~
{\sg_i\over \Mp} \simeq \left(H_1\over \Mp\right)^{1/6} z_d^{1/4} z_\sg^{-7/12},
\label{c7}
\eeq
provide strong constraints on the possible spectral distribution. 

There are further important constraints arising from the self-consistency of the considered pre-big bang scenario, and from its consistency with standard cosmological phenomenology. In particular: the axion decay scale $H_d$ must be larger than the scale $H_N \sim (1 {\rm Mev})^2/\Mp$ of standard nucleosynthesis. Also, the typical frequency scale of $f_{\rm LSS}$ of Large Scale Structures, related to the {\it pivot} frequency scale $f_* \simeq 0.05 \,{\rm Mpc}^{-1}$ by $f_{\rm LSS} \simeq 60 f_*$, must be smaller than the spectral scale $f_s$ marking the beginning of the string phase: namely, $f_* < f_{\rm LSS} \laq f_s$. Finally, the normalisation of the scalar spectrum at the {\it pivot} scale, written in terms of the observed scalar spectral index $n_s \simeq 0.965$ and of the scalar amplitude $P_s(f_*) \simeq 2.1 \times 10^{-9}$, gives a phenomenological condition which automatically fixes $H_1$ in terms of all the other spectral parameters. 

All the above-mentioned constraints have been explicitly written (in a convenient logarithmic form) and applied to the parameters of the GW spectrum (\ref{c1}), for a general (dual and non-dual) non-minimal scenario,  in previous papers (see in particular \cite{1} for an updated discussion). The resulting allowed regions of the spectral parameters has been used in this paper to select the explicit examples of non-minimal models and the related spectra presented in Sects. \ref{sec31}, \ref{sec32}.


\end{document}